\documentclass[conference]{IEEEtran}
\IEEEoverridecommandlockouts
\usepackage{citesort}
\usepackage{amsmath,amssymb,amsfonts}
\usepackage{algorithmic}
\usepackage{graphicx}
\usepackage{textcomp}
\usepackage{url}
\usepackage{mathtools}
\usepackage{caption}
\usepackage{subcaption}
\DeclarePairedDelimiter{\ceil}{\lceil}{\rceil}
\def\BibTeX{{\rm B\kern-.05em{\sc i\kern-.025em b}\kern-.08em
    T\kern-.1667em\lower.7ex\hbox{E}\kern-.125emX}}

\DeclareMathOperator*{\argmax}{arg\,max}

\begin{document}

\title{Visible Light Communication for Next Generation Untethered Virtual Reality Systems
\thanks{This work has been supported in part by NSF Awards CCF-1528030, ECCS-1711592, CNS-1836909, and CNS-1821875.}
}



	\author{\IEEEauthorblockN{Mahmudur Khan and Jacob Chakareski}
		\IEEEauthorblockA{Laboratory for Virtual and Augmented Reality Immersive Communications\\The University of Alabama,
				Tuscaloosa, AL 35487}
		Email: \{mkhan6, jacob\}@ua.edu. Web: http://lion.ua.edu
	}

\maketitle

\begin{abstract}
Virtual and augmented reality (VR/AR) systems are emerging technologies requiring data rates of multiple Gbps. Existing high quality VR headsets require connections through HDMI cables to a computer rendering rich graphic contents to meet the extremely high data transfer rate requirement.
Such a cable connection limits the VR user's mobility and interferes with the VR experience. Current wireless technologies such as WiFi cannot support the multi-Gbps graphics data transfer. Instead, we propose to use visible light communication (VLC) for establishing high speed wireless links between a rendering computer and a VR headset. But, VLC transceivers are highly directional with narrow beams and require constant maintenance of line-of-sight (LOS) alignment between the transmitter and the receiver. Thus, we present a novel multi-detector hemispherical VR headset design to tackle the beam misalignment problem caused by the VR user's random head orientation. We provide detailed analysis on how the number of detectors on the headset can be minimized while maintaining the required beam alignment and providing high quality VR experience.
\end{abstract}


\section{Introduction}
\label{sec:intro}
VR/AR systems have become exceedingly popular recently by enabling new immersive digital experiences. The application areas of VR/AR is not only limited to gaming and entertainment, but also education and training, environmental and weather sciences, disaster relief, and healthcare \cite{chakareski2018viewport,apostolopoulos2012road}. The tremendous potential of VR/AR systems has resulted in accelerated market growth and is expected to reach $\$120$ billion by the year 2022 \cite{tmerel2017jan}. However, high quality VR systems like Oculus Rift \cite{occulus2018rift} and HTC Vive \cite{htc2018vive} render the rich graphics content on a powerful computer. Such systems require HDMI and USB cable connections between the computer and the VR headset for transferring the multi-Gbps data. These cables limit the VR user's mobility, deteriorate his quality of experience, and present potential tripping/neck injury hazards. Standalone VR systems like Samsung Gear \cite{samsung2018gear} or Google Daydream \cite{google2018daydream} render the graphics contents on the VR headset or on a smart phone. Although these systems are wireless, the rendering quality is limited by the capability of the headset or the phone. Existing wireless technologies, such as WiFi, cannot provide the high data rates required by high quality VR systems. Streaming 360$^\circ$ VR videos over them is even more constraining, due to the even higher data rate demand of such real remote scene multi-perspective immersion content \cite{Chakareski:17b,ChakareskiVS:12,CorbillonDSC:17a}. 

Hence, academic and industry researchers are considering alternative wireless technologies, e.g., mmWave, to support multi-Gbps data rates. However, \cite{cuervo2018creating} estimates that in the near future, the VR headsets will render life-like experiences and the required data transfer rate will exceed 1 Tbps. Supporting such high throughput wireless links will be extremely challenging even for mmWave systems. Thus, we explore the use of VLC to enable high speed wireless VR systems.

\begin{figure}
	\centering
	\includegraphics[width=\linewidth]{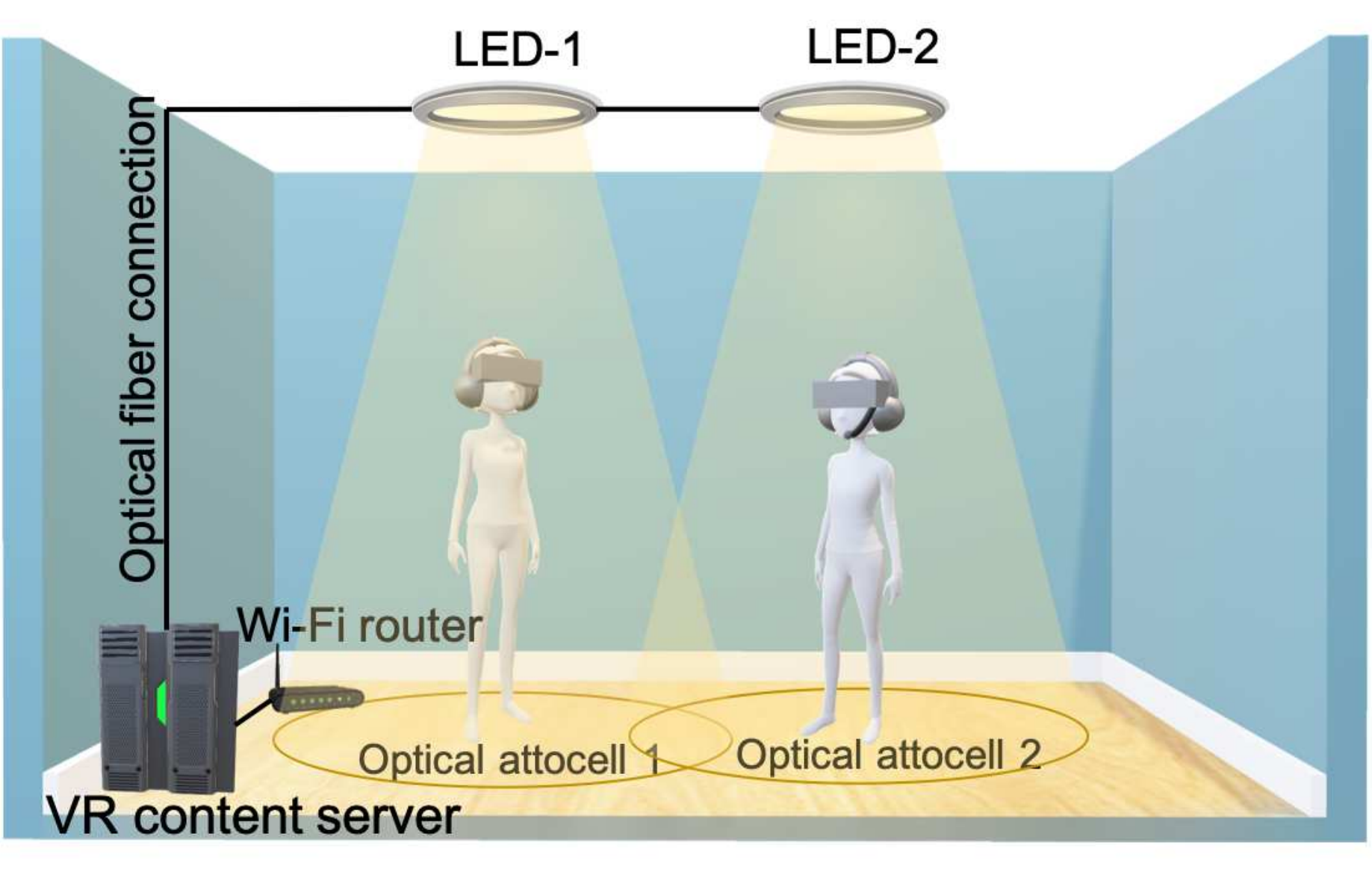}
	\caption{VLC-enabled VR-arena. Pointed optical transmitters provide high-quality 360$^\circ$ VR videos. Omnidirectional Wi-Fi link provides baseline quality video and control channel.}
	\label{fig:VLC-VR-Arena}
	\vspace{-10pt}
\end{figure}

Optical wireless communication (OWC) has the potential to provide optical-level wireless communication speeds.  It uses the unlicensed optical spectrum and mostly uses the same basic optoelectronic technology as fiber optic communications. A highly useful feature of OWC is its inherent signal security due to the containment of optical wireless signals behind walls and their low probability of interception and detection \cite{zhou2011enlargement}. Also, the high directionality of OWC provides high spatial reuse and larger network capacity. VLC is a subset of the OWC technology that operates in the visible light spectrum band between 400-800 THz. In the recent years, VLC has attracted significant research interest due to the development of light-emitting-diode or LED technology which can provide both illumination and high speed wireless data transfer simultaneously \cite{komine2004fundamental}. LEDs are anticipated to dominate the illumination market by 2020 \cite{baumgartner2012lighting}. Since VLC allows both illumination and communication at the same time, it has attracted several indoor and outdoor applications such as lighting infrastructure in buildings with building-to-building wireless communication, high speed data transfer between autonomous cars, high speed communication inside offices, houses, airplane cabins, and even in VR arenas. However, due to the high directionality, VLC requires line-of-sight (LOS) alignment between transmitter and receiver beams, which becomes challenging especially in mobile scenarios.

In this paper, we present an optimized headset design for a VLC based VR system. We consider an indoor VR-arena (Fig. \ref{fig:VLC-VR-Arena}) where multiple LED transmitters are deployed to create a network of multiple small cells, a.k.a. optical attocells \cite{chen2016downlink}. This network provides simultaneous high speed wireless connectivity to multiple VR-users situated at any location inside the arena. Moreover, such a network of multiple cells can help maintain wireless connectivity when a VR-user is in motion. We also consider that the VR-headset has a hemispherical helmet/cap and is equipped with multiple photo-detectors (PDs) placed over the cap's surface (Fig. \ref{fig:VLC-VR-Headset}). The proposed multi-detector design helps maintain wireless connectivity during random headset orientation caused by the VR-user's head movement. We provide detailed analysis on how choosing the appropriate number of PDs with proper field-of-view (FOV) can provide both full link connectivity and a high quality VR experience.

The rest of the paper is organized as follows. In Section \ref{sec:relatedwork}, we discuss the literature on both VLC and VR. In Section \ref{sec:VLC-VR}, we present the details of the proposed VLC-enabled VR system. We describe the VLC-channel model in Section \ref{sec:VLC-channel} and the evaluation of the proposed system in Section \ref{sec:results}. Finally, we conclude in Section \ref{sec:conclusion}.

\section{Related Work}
\label{sec:relatedwork}
Here, we first review the state-of-the-art of VLC research. Then, we overview existing work on wireless VR systems.
\subsection{Visible light communication}
\label{sec:vlc}
VLC systems typically use white LEDs as transmitters. The modulation bandwidth of such LEDs is a few MHz which limits the transmission rate of a VLC system \cite{ghassemlooy2012optical}. To achieve the desired high speed data rate, different methods have been proposed, such as, frequency domain equalization \cite{le2009100}, using optical filters, modulation techniques like orthogonal frequency division multiplexing \cite{hashemi2008orthogonal,chen2017non}, multiple-input multiple output (MIMO) \cite{zeng2009high,butala2013svd}, and using red-green-blue (RGB) LEDs \cite{kottke20121}. In \cite{cossu20123}, a data rate of 3.4 Gbps was reported which was achieved using wavelength division multiplexing (WDM) and RGB LEDs. In \cite{hussein201510,hussein201625}, RGB laser diodes (RGB-LDs) were used instead of LEDs, since LDs can provide the high modulation bandwidth of multiple GHz. It has been shown in \cite{neumann2011four,denault2013efficient} that a combination of visible RGB lasers along with a diffuser can generate white light that has comparable properties with white LED sources. \cite{hussein201625} has shown that data rates of as high as 25Gbps can be achieved using RGB-LDs.

An indoor VLC network has multiple attocells to provide full wireless communication coverage. Inter-cell-interference (ICI) becomes an issue in such networks due to the overlapping of optical transmission beams. Different approaches have been proposed to mitigate ICI, for example, frequency division based carrier allocation \cite{kim2012mitigation}, a dynamic interference-constrained sub-carrier reuse algorithm \cite{bykhovsky2014multiple}, and using angle diversity receivers (ADRs). In \cite{chen2014angle}, an ADR was proposed with a hemispherical base where the PDs were placed in one or two layers. In \cite{chen2018reduction,nuwanpriya2015indoor}, different ADR designs were presented where the PDs are placed at a given inclination angle on the same horizontal plane. Such ADRs can help maintain a VLC link in stationary scenario or during yaw movement but not in the presence of roll and/or pitch movement of the receiver. In this paper, we present a multi-detector VR-headset design that can provide connectivity for any orientation of the VR-user's head, such as, roll, pitch, and yaw (Fig. \ref{fig:roll}).

\subsection{Wireless virtual reality}
\label{sec:wvr}
Existing wireless VR systems like Samsung Gear or Google Cardboard rely on smart phones and cannot process rich graphics content. VR systems like Zotac require the user to carry a processor unit in backpack. High quality VR systems like Occulus Rift and HTC Vive are not wireless and requires HDMI and USB cable connection to a computer for rendering the rich graphic content. For providing wireless VR with high quality, 60 GHz communication is being explored as a possible solution. In \cite{zhong2017building}, a proof-of-concept VR system with programming capability on the headset is proposed that uses WiGig modules for wireless connection. The authors in \cite{abari2017enabling} developed a mmWave reflector that helps maintain connection with a VR-headset in the event of blockage. In \cite{rahman2018fso}, the authors proposed using Free-space-optics (FSO), a.k.a. OWC to replace the HDMI cable on a VR-headset. They proposed a mechanical steering mechanism for both a ceiling mounted transmitter and a receiver on the headset to maintain line-of-sight alignment.

In this paper, we propose a VLC based VR systems where multiple attocells are used to maintain optical wireless links during user mobility or in presence of multiple VR-users. We also present a VR-headset design that leverages from multiple highly directional PDs to maintain beam alignment when the VR-user moves his/her head in random orientations.

\section{VLC-enabled VR system}
\label{sec:VLC-VR}
\begin{figure}
	\centering
	\includegraphics[width=\linewidth]{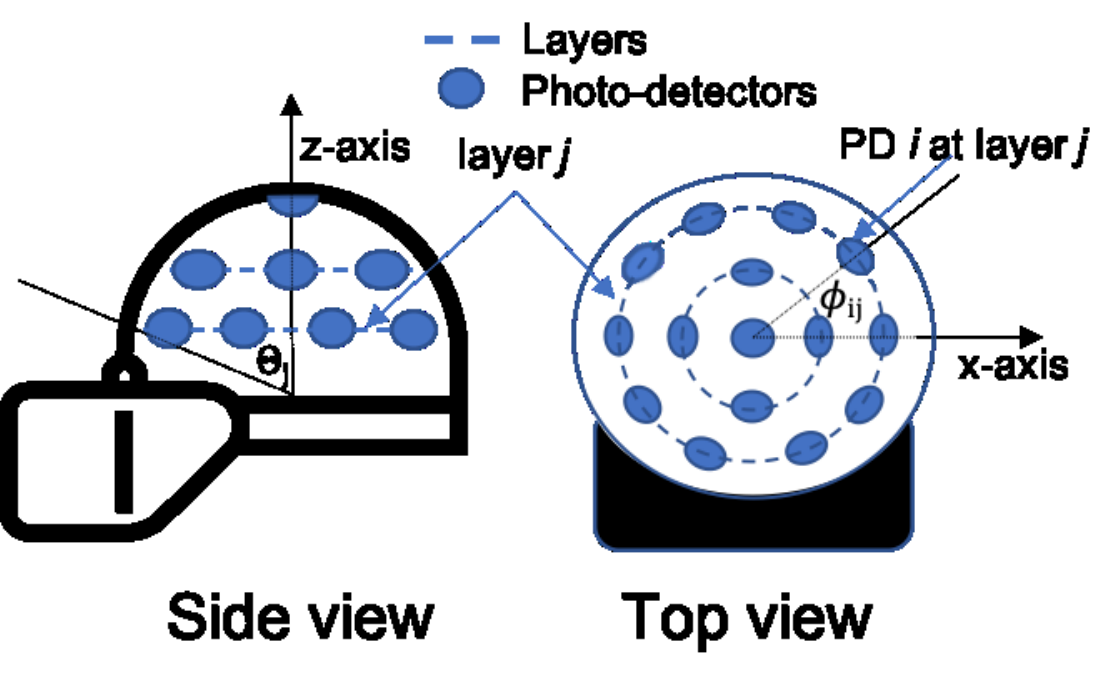}	
	\caption{VR-headset equipped with photo-detectors}
	\label{fig:VLC-VR-Headset}
	\vspace{-10pt}	
\end{figure}

\subsection{Optical attocell VR-arena network}
We consider the VR-arena to be equipped with multiple LEDs or LDs as optical transmitters which are connected to a VR content server through fiber optic cables (Fig. \ref{fig:VLC-VR-Arena}). Each optical transmitter creates a small cell or an optical attocell. Thus, an attocell network is established that helps provide coverage in the whole VR-arena both in terms of illumination and high speed data transfer. Each VR-user is served by one of the transmitters and multiple users can be served by the same transmitter simultaneously. A user is assigned a transmitter from which it has the shortest distance. Existing systems like the HTC Vive and Occulus Rift can track the position and head-orientation of the user. As the user moves within the arena he/she can be switched to other transmitters for service depending on his/her location.

\subsection{VR-user headset orientation}
\label{realdata}
\begin{figure}
	\begin{subfigure}{.5\linewidth}
	\centering%
	\includegraphics[width=.9\linewidth]{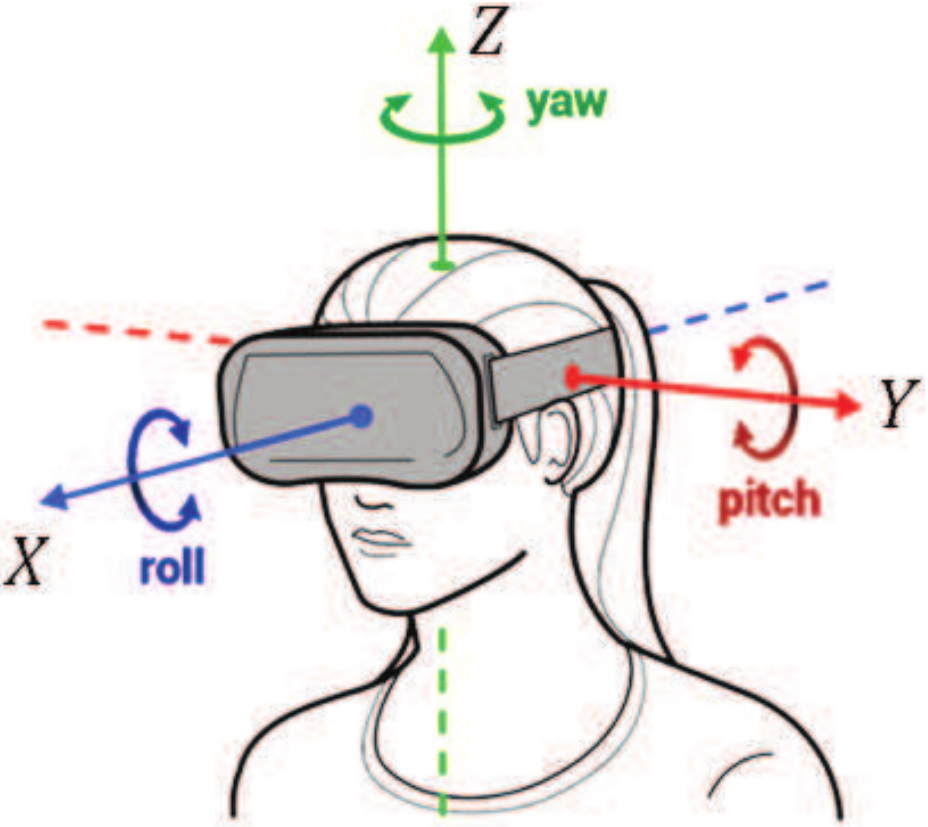}
	\subcaption{Yaw, pitch, roll movements}
	\label{fig:roll}
	\end{subfigure}%
	\begin{subfigure}{.5\linewidth}	
	\centering
	\includegraphics[width=\linewidth]{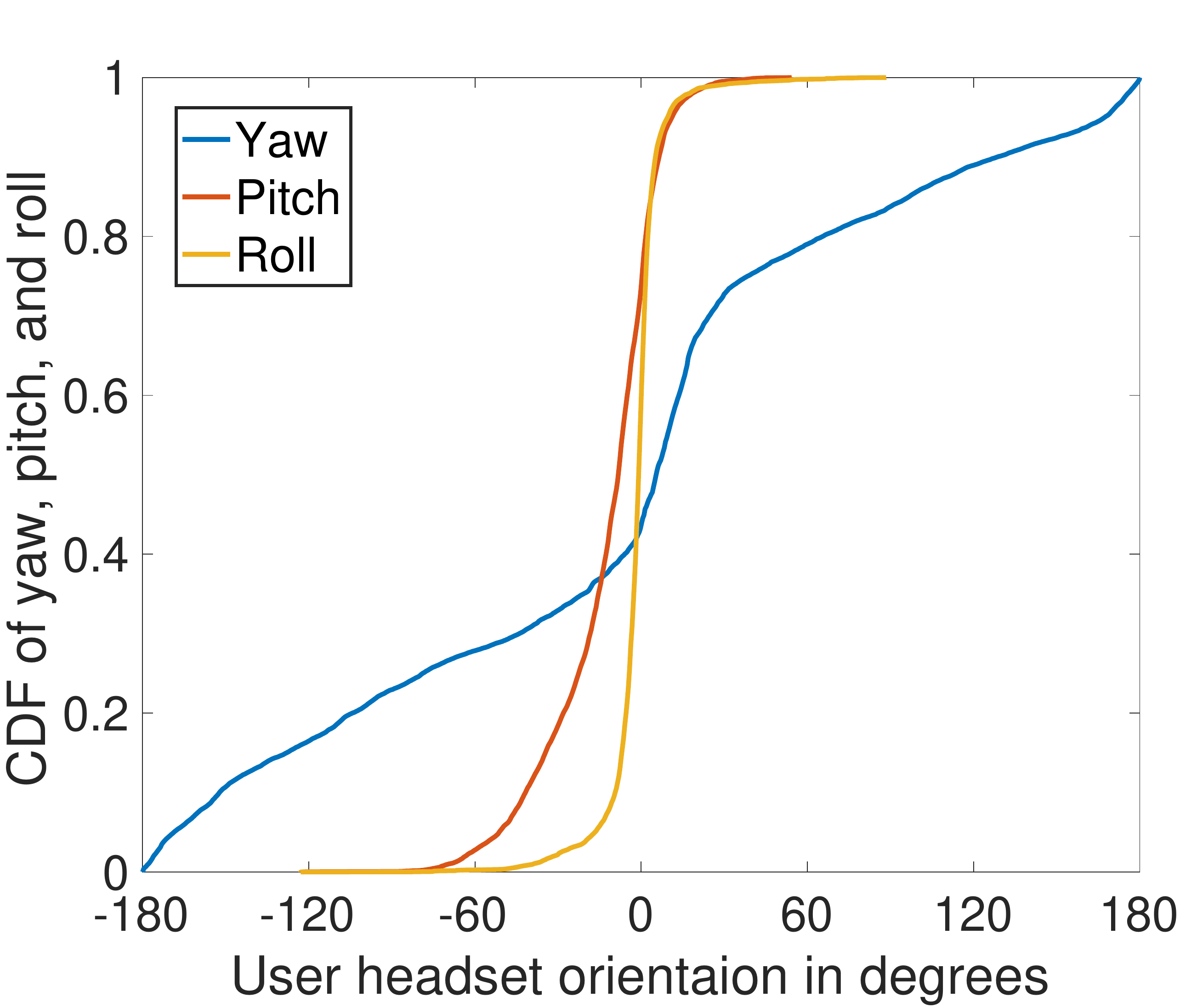}
	\subcaption{Head orientation data}
	\label{fig:real_orientation_data}
	\end{subfigure}%
	\caption{User head-orientation while watching 360$^o$ VR videos}
\end{figure}

Fig. \ref{fig:real_orientation_data} displays the cumulative distribution of yaw, pitch, and roll (Fig. \ref{fig:roll}) values of user head movement collected from 40 users watching 6 VR 360$^o$ videos \cite{corbillon2017360}. We observed that the pitch and roll values stay within -55$^o$ to 55$^o$ in 96.13$\%$ and 99.43$\%$ of the cases respectively. The yaw values can be anywhere between -180$^o$ and $180^o$.

\subsection{Multi-detector VR-headset}
We propose a hemispherical headset equipped with multiple PDs to receive the optical signal from the LED/LD transmitters. The availability of multiple PDs helps in maintaining optical links for any orientation of the user's headset. We consider optical wireless links for download, and Wifi for feedback and upload activity. We also assume that the Wifi link continuously provides baseline quality VR videos for seamless connectivity. Moreover, the receiver headset will require a processing unit to decode the received rich graphic content. The decoding can also be performed by offloading the task to a nearby edge computer. Designing an efficient processing unit at the receiver side for decoding high quality VR-content requires further exploration and is out-of-scope of this work. We focus on designing a VLC-enabled VR-headset that can maintain optical links for any user head movement.

We present a VR-headset design as shown in Fig. \ref{fig:VLC-VR-Headset}, where one PD is placed on top and the rest of the PDs are placed at different hemispherical layers on the headset. Considering circular shaped PDs of the same radius $r_{PD}$, the minimum angular distance between two such PDs is $\Theta_{min} = 2\times\tan^{-1}(\frac{r_{PD}}{r_{Headset}})$, where $r_{Headset}$ is the radius of the VR-headset hemisphere. Now, for an angular distance $\Theta_d\ge \Theta_{min}$, the total number of PD layers $N_L = \ceil[\big]{\frac{90-\Theta_d/2}{\Theta_d}}$. The angular distance between two such layers is:

\begin{equation}
\Theta_z = \begin{cases}
\frac{90^o}{N_L}, & \text{if $\Theta_d\times N_L>90^o$} \\
\Theta_d, & \text{otherwise}.
\end{cases}
\end{equation}

\noindent The top PD is placed at an inclination angle of 0$^o$ and azimuth angle of 0$^o$. A PD placed at the layer $\emph j$ has an inclination angle of $\Theta_j = \Theta_z\times j$ and the number of PDs in that layer is:

\begin{equation}
N_j = \ceil[\big]{\frac{360^o}{\Theta_d}\times\sin(\Theta_j)}, j = 1, 2, ..., N_L
\end{equation}

\noindent The azimuth angle of the $i$-th detector on layer $j$ is given by:

\begin{equation}
\Phi_{ij} = \frac{360^o}{N_j}\times (i-1), i = 1, 2, ..., N_j
\end{equation}

\noindent So, the total number of PDs on the headset for an angular distance of $\theta_d$ can be determined as,

\begin{equation}
N_{PD} = 1 + \sum_{j=1}^{N_L} N_j
\label{eq:Npd}
\end{equation}

\noindent Let us consider that the coordinates of the top PD in vertical orientation is $(x_0, y_0, z_0)$. Then, the coordinates of the $i$-th PD at layer $j$ is $(x_{ij}, y_{ij}, z_{ij})$. Here,

\begin{equation}
\begin{split}
x_{ij} &= x_0 + r_{Headset}\cos\Theta_j\sin\Phi_{ij} \\
y_{ij} &= y_0 + r_{Headset}\sin\Theta_j\sin\Phi_{ij} \\
z_{ij} &= z_0 - r_{Headset}\cos\Phi_{ij}
\end{split}
\end{equation}

\noindent Now, if the headset's orientation changes by $R_z R_y R_x$, the coordinates of the PD will be:

\begin{equation}
\resizebox{.9\hsize}{!}{$
\begin{split}
\begin{bmatrix}
x'_{ij}\\y'_{ij}\\z'_{ij}
\end{bmatrix} =  R_zR_yR_x\begin{bmatrix}
x_{ij}\\y_{ij}\\z_{ij}
\end{bmatrix}, \text{ where, }
R_x = \begin{bmatrix}
1 & 0 & 0\\
0 & c_x & -s_x\\
0 & s_x & c_x
\end{bmatrix}, \\
R_y = \begin{bmatrix}
c_y & 0 & s_y\\
0 & 1 & 0\\
-s_y & 0 & c_y
\end{bmatrix}\text{, }
R_z = \begin{bmatrix}
c_z & -s_z & 0\\
s_z & c_z & 0\\
0 & 0 & 1
\end{bmatrix}
\end{split}$}
\label{eq:rmatrix}
\end{equation}

\noindent Here, $R_x$, $R_y$, and $R_z$ are rotation matrices representing $roll$, $pitch$, and $yaw$ respectively. The rotation angles around the $x$, $y$, and $z$ axes are $\theta_x$, $\theta_y$, and $\theta_z$ respectively. Also, $c_x = \cos\theta_x$, $c_y = \cos\theta_y$, $c_z = \cos\theta_z$, $s_x = \sin\theta_x$, $s_y = \sin\theta_y$, and $s_z = \sin\theta_z$.

\section{VLC Channel Model}
\label{sec:VLC-channel}

\begin{figure}
	\centering
	\includegraphics[width=.75\linewidth]{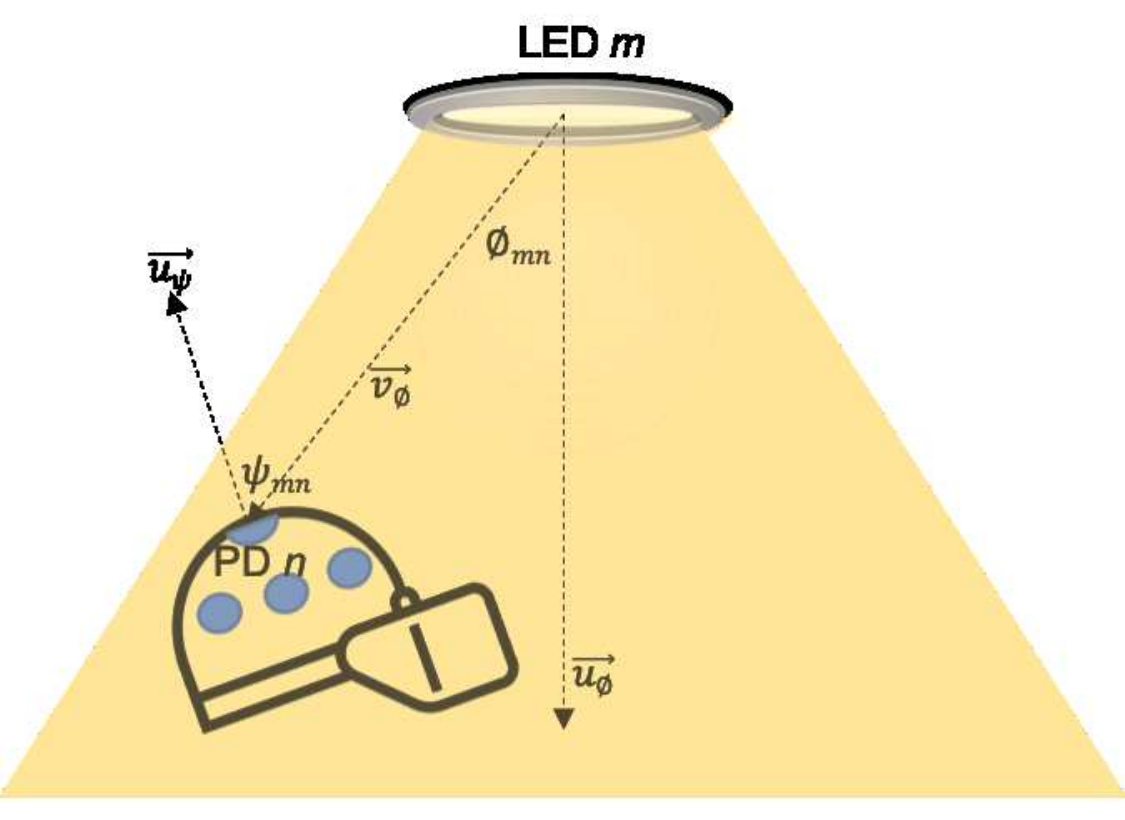}
	\vspace{-5 pt}	
	\caption{Irradiance and incident angle between an LED and PD.}
	\label{fig:angles}
	\vspace{-10pt}	
\end{figure}

\subsection{Received signal power}
We consider a line-of-sight (LOS) optical channel and assume that the LED lighting has a Lambertian radiation pattern \cite{ghassemlooy2012optical}. Now, if the number of LED/LD transmitters in the VR-arena is $N_{LED}$ and if a headset has $N_{PD}$ photo-detectors, then the received power $P_{mn}^{r}$ at the $n$-th ($n = 1, 2, ..., N_{PD}$) PD from the $m$-th ($m = 1, 2, ..., N_{LED}$) transmitter is:
\begin{equation}
\resizebox{.9\hsize}{!}{$P_{mn}^{r}   =  P_m^{t}\frac{(m_l + 1)A_{PD}}{2\pi d_{mn}^2}\cos^{m_l}(\phi_{mn})T_s(\psi_{mn})g(\psi_{mn})\cos(\psi_{mn})$}
\end{equation}
\vspace{-10pt}	

\noindent where $P_n^{t}$ is the transmitted optical power of an LED, $m_l = log(2)/log(cos\Psi)$ is the order of Lambertian emission with $\Psi$ being the LED divergence angle, $A_{PD}$ is the area of a PD, $d_{mn}$ is the distance between LED $\emph m$ and PD $\emph n$, $\phi_{mn}$ is the irradiance angle, $\psi$, is the incident angle, and $T_s(\psi)$ is the filter transmission. Here, $g(\psi)$ and $\beta$ are the optical concentrator gain and FOV respectively, and $g(\psi_{mn}) = \eta^2/sin^2(\beta)$, where $\eta$ is the refractive index. Now, if the $m$-th LED's coordinates are ($x_m, y_m, z_m$) and the $n$-th PD's coordinates are ($x_n,y_n, z_n$), the irradiance angle is given by (Fig. \ref{fig:angles}):

\vspace{-5pt}	
\begin{equation}
\phi_{mn} = \tan^{-1}\frac{||\overrightarrow{u_{\phi}}\times \overrightarrow{v_{\phi}}||}{\overrightarrow{u_{\phi}}.\overrightarrow{v_{\phi}}}
\end{equation}

\noindent where, $\overrightarrow{u_{\phi}} = [0, 0, z_m]$, and $\overrightarrow{v_{\phi}} = [x_m-x_n, y_m-y_n, z_m-z_n]$. And, the incident angle is given by:

\vspace{-5pt}	
\begin{equation}
\psi_{mn} = \tan^{-1}\frac{||\overrightarrow{u_{\psi}}\times \overrightarrow{v_{\psi}}||}{\overrightarrow{u_{\psi}}.\overrightarrow{v_{\psi}}}
\end{equation}

\noindent where, $\overrightarrow{u_{\psi}} = [\cos\Theta_m'\sin\Phi_m', \sin(\Theta_m')\sin\Phi_m', \cos\Phi_m']$, and $\overrightarrow{v_{\psi}} = [x_n-x_m, y_n-y_m, z_n-z_m]$. Here, $\Phi_m'$ and $\Theta_m'$ represent the inclination angle and azimuth angle, respectively, of a PD at a given orientation.

The received signal at a PD includes shot noise and thermal noise. The variances of these noises are:

\begin{equation}
\sigma_{n,shot}^2 = 2qR_{PD}\sum_{m=1}^{N_{LED}}P_{mn}^rB + 2qI_bI_2B
\end{equation}


\begin{equation}
\resizebox{.85\hsize}{!}{$\sigma_{n,thermal}^2 = \frac{8\pi\kappa T_k}{G}C_{PD}A_{PD}I_2B^2 + \frac{16\pi^2\kappa T_k \Gamma}{g_m}C_{PD}^2A_{PD}^2I_3B^3$}
\end{equation}

\noindent where, $q$ is electron charge, $R_{PD}$ is PD responsivity, $B$ is the electrical filter bandwidth, $I_b$ is the photocurrent due to background radiation, $I_2 = 0.562$, $\kappa$ is Boltzmann's constant, $T_k$ is absolute temperature, $G$ is the open-loop voltage gain, $C_{PD}$ is the fixed capacitance of PD per unit area, $A_{PD}$ is the PD area, and $I_3 = 0.0868$. So, total noise at a PD, $\sigma_{n}^2 = \sigma_{n,shot}^2 + \sigma_{n,thermal}^2$ \cite{komine2004fundamental}.

\subsection{Signal-to-interference-and-noise-ratio (SINR)}
As mentioned earlier, an attocell network is established in the VR-arena to provide full communication coverage. The overlapping of transmitter beams in such a multi-cell network causes inter cell interference. When a receiver is located at the overlapping area of multiple transmitter beams, it may receive signal not only from the transmitter it is assigned to, but also from other transmitters. The signals received from these ``other" transmitters are regarded as interference.

The proposed VR-headset design consists of multiple PDs. So, signals received by all the PDs on a headset are combined to produce the resultant signal. We consider three diversity combining techniques, equal-gain-combining (EGC), select-gain-combining (SBC), and maximal-ratio-combining (MRC) \cite{chen2014angle,chen2018reduction,brennan2003linear}.

In EGC, the signals received from all the PDs are simply added with equal weights and the resulting SINR can be determined as:

\begin{equation}
\resizebox{.85\hsize}{!}{$SINR_m^{EGC} = \frac{\left(\sum_{n=1}^{N_{PD}}R_{PD}P_{mn}^{r}\right)^2}{\sum_{n=1}^{N_{PD}}\left(\sum_{m'=1,m'\ne m}^{N_{LED}}(R_{PD}P_{mn}^{r})^2 + \sigma_n^2\right)}$}
\end{equation}

In SBC, the SINR at each PD is calculated separately as:
\begin{equation}
\resizebox{.85\hsize}{!}{$SINR_{mn} = \frac{(R_{PD}P_{mn}^{r})^2}{\sum_{n=1}^{N_{PD}}\left(\sum_{m'=1,m'\ne m}^{N_{LED}}(R_{PD}P_{mn}^{r})^2 + \sigma_n^2\right)}$}
\end{equation}

\noindent And, the output of the PD with the highest SINR is chosen as the final signal. So, the SINR for the whole receiver can be determined as: $SINR_m^{SBC} = \max (SINR_{mn})$.


Finally, in MRC, the output signal of each PD is multiplied by a weight $\lambda_{mn} = SINR_{mn}$. The SINR for the MRC technique can be presented by:
\begin{equation}
\resizebox{.85\hsize}{!}{$SINR_m^{MRC} = \frac{\left(\sum_{n=1}^{N_{PD}}\lambda_{mn}R_{PD}P_{mn}^{r}\right)^2}{\sum_{n=1}^{N_{PD}}\left(\sum_{m'=1,m'\ne m}^{N_{LED}}(\lambda_{mn}R_{PD}P_{mn}^{r})^2+\lambda_{mn}^2\sigma_n^2\right)}$}
\end{equation}

\section{Simulation Results and Analysis}
\label{sec:results}
We performed MATLAB simulations to optimize the VR-headset design and evaluate its performance in a VLC-enabled VR-arena with dimensions of $5m\times5m\times3m$. We consider the VR-user's height to be $1.33m$. The VR-arena is equipped with four LED/LD transmitters placed at coordinates (1.25, 1.25, 3), (1.25, 3.75, 3), (3.75, 1.25, 3), and (3.75, 3.75, 3), considering that the units of coordinates are in meters. Table \ref{tab:tab1} provides the rest of the simulation parameters.

\subsection{Ensuring full connectivity}
We consider that all the PDs have the same half-angle FOV $\beta^o$. For a given angular distance between two PDs $\Theta_d$, the FOV should be chosen in such a way that an optical link between an LED/LD transmitter and the headset receiver is maintained for any orientation of the user's head. Let us assume that a headset can be oriented in $N_{orientaion}$ ways through different combination of yaw, pitch, and roll values. Now, if the optical link between a transmitter and a headset is active for $N_{up}^{\alpha}$ different orientations, the connectivity provided by that headset is $cvg^{\alpha} = (N_{up}^{\alpha}/N_{orientaion})\times 100\%$, where $\alpha = 2\beta/\Theta_d$. If the $\alpha_{min}$ is the minimum value of $\alpha$ for which $cvg^{\alpha}=100\%$, then, $\alpha \ge \alpha_{min}$.

\begin{figure}
	\centering
	\includegraphics[width=.75\linewidth]{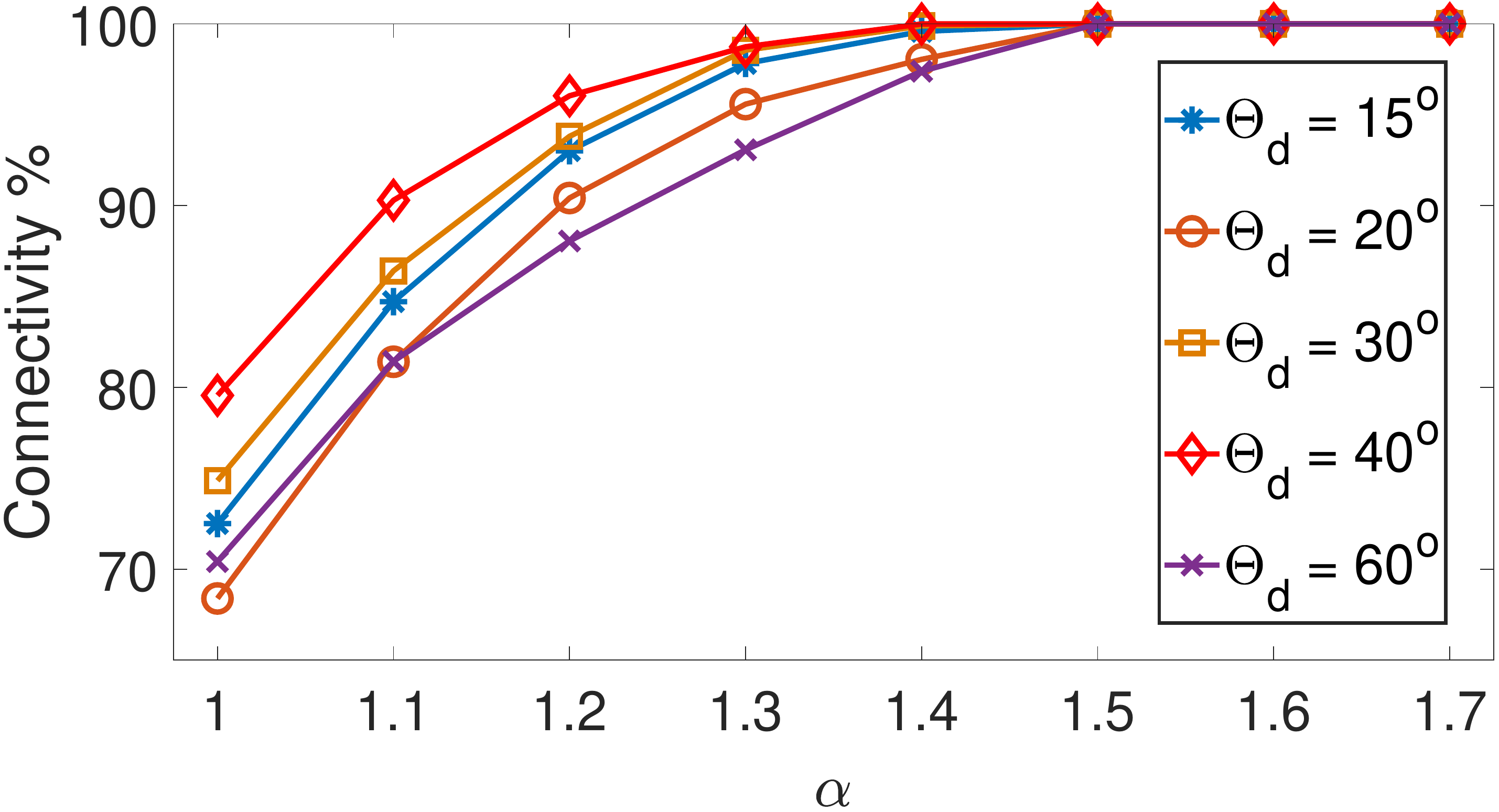}
	\vspace{-5pt}
	\caption{Required $\alpha\ge 1.5$ for full VLC connectivity}
	\label{fig:find_alpha_min}
	\vspace{-15pt}
\end{figure}

We first conducted simulations by considering a VR-user located at position (1.25, 1.25, 1.33) and LED-1 was the corresponding transmitter. We changed the VR-headset's roll and pitch values between $-90^o$ and $+90^o$, and yaw values between $0^o$ and $+359^o$ in increments of $1^o$. So, $N_{orientaion} = 181\times181\times360$. The FOV was calculated from the relation $\alpha = FOV/\Theta_d$ or $2\beta/\Theta_d$. Fig. \ref{fig:find_alpha_min} displays the percentage of time an optical link is active between the transmitter and the receiver for $N_{orientaion}$ different orientations of the headset. We can observe that for almost all different angular distances ($\Theta_d = 15^o$, $20^o$, $30^o$, $40^o$, $60^o$) between the PDs on the headset, $100\%$ connectivity is achieved for any orientation when $\alpha\ge1.5$. So, $\alpha_{min}=1.5$ and thus, $FOV\ge1.5\times\Theta_d$.

\begin{table}
	\caption{Simulation Parameters}
	\begin{center}
		\begin{tabular}{|c|c|}
			\hline
			VR-headset radius, $r_{Headset}$ & 7.62cm \\
			\hline
			Divergence angle of an LED & $60^o$ \\
			\hline
			Transmission power of an LED, $P^t$ & 10W \\
			\hline
			Refractive index of optical concentrator, $\eta$ & 1.5 \\
			\hline
			Gain of optical filter, $g(\psi)$ & 0.9 \\
			\hline
			PD radius, $r_{PD}$ & 2.5mm \\
			\hline
			PD responsivity, $R_{PD}$ & 0.53A/W \\
			\hline
			Modulation bandwidth, B & 10MHz \\
			\hline
			Background current, $I_b$ & 5100$\mu A$ \\	
			\hline
		\end{tabular}
		\label{tab:tab1}
	\end{center}
	\vspace{-15pt}
\end{table}

\subsection{Optimum value of $\alpha$}
The value of $\alpha$ should be chosen in such a way that not only full connectivity between a transmitter and a receiver is achieved but also a high SINR is maintained. Thus, the desired value of $\alpha$ should be, $\alpha^{opt} = \argmax_{\alpha}SINR(\alpha)$.

\begin{figure}
	\centering
	\begin{subfigure}{.5\linewidth}
		\centering%
		\includegraphics[width=.75\linewidth]{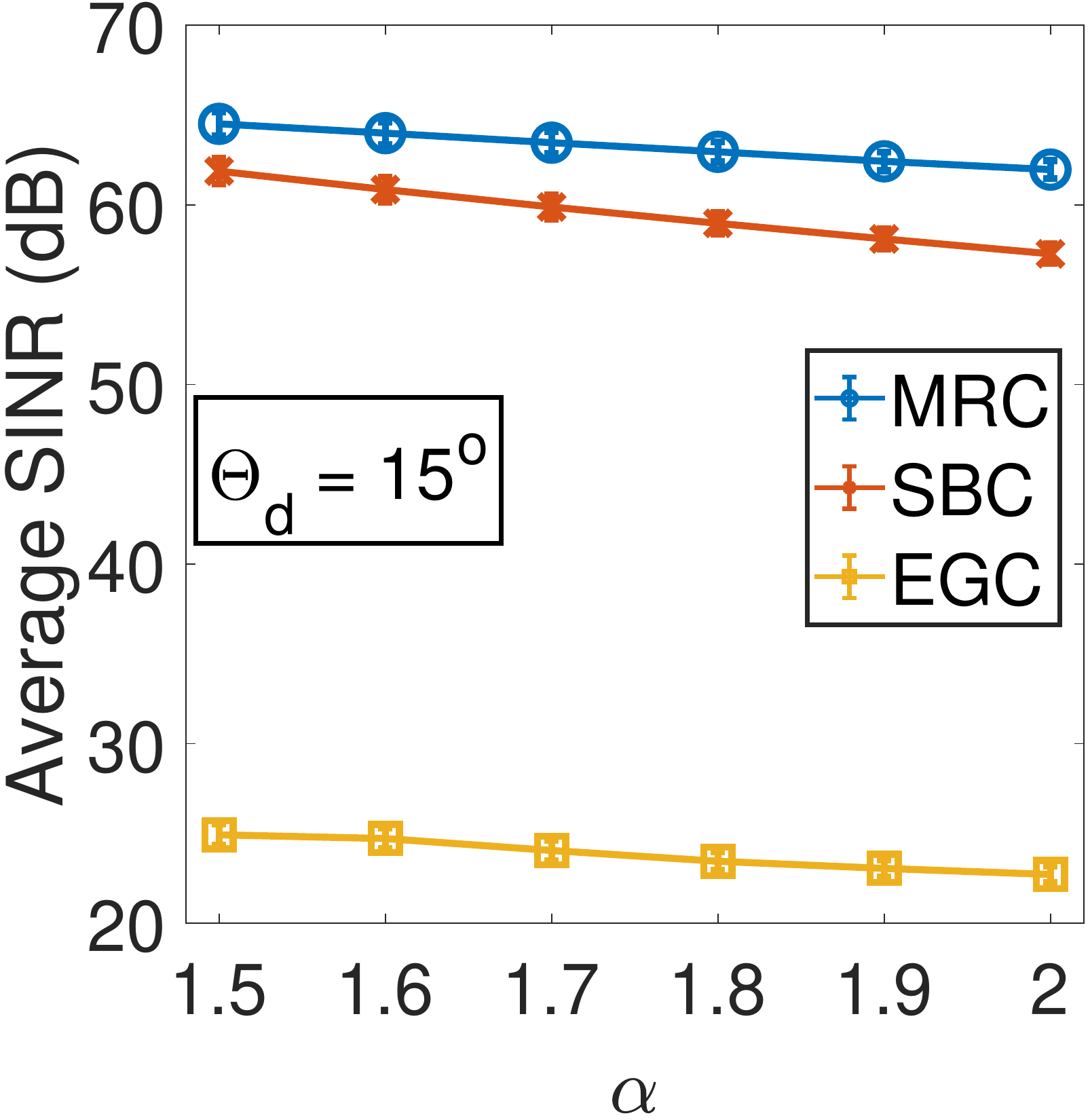}
		\vspace{-5pt}
		\subcaption{$\Theta_d = 15^o$}
		\label{fig:sinr_AL_15}
	\end{subfigure}%
	\begin{subfigure}{.5\linewidth}
		\centering%
		\includegraphics[width=.75\linewidth]{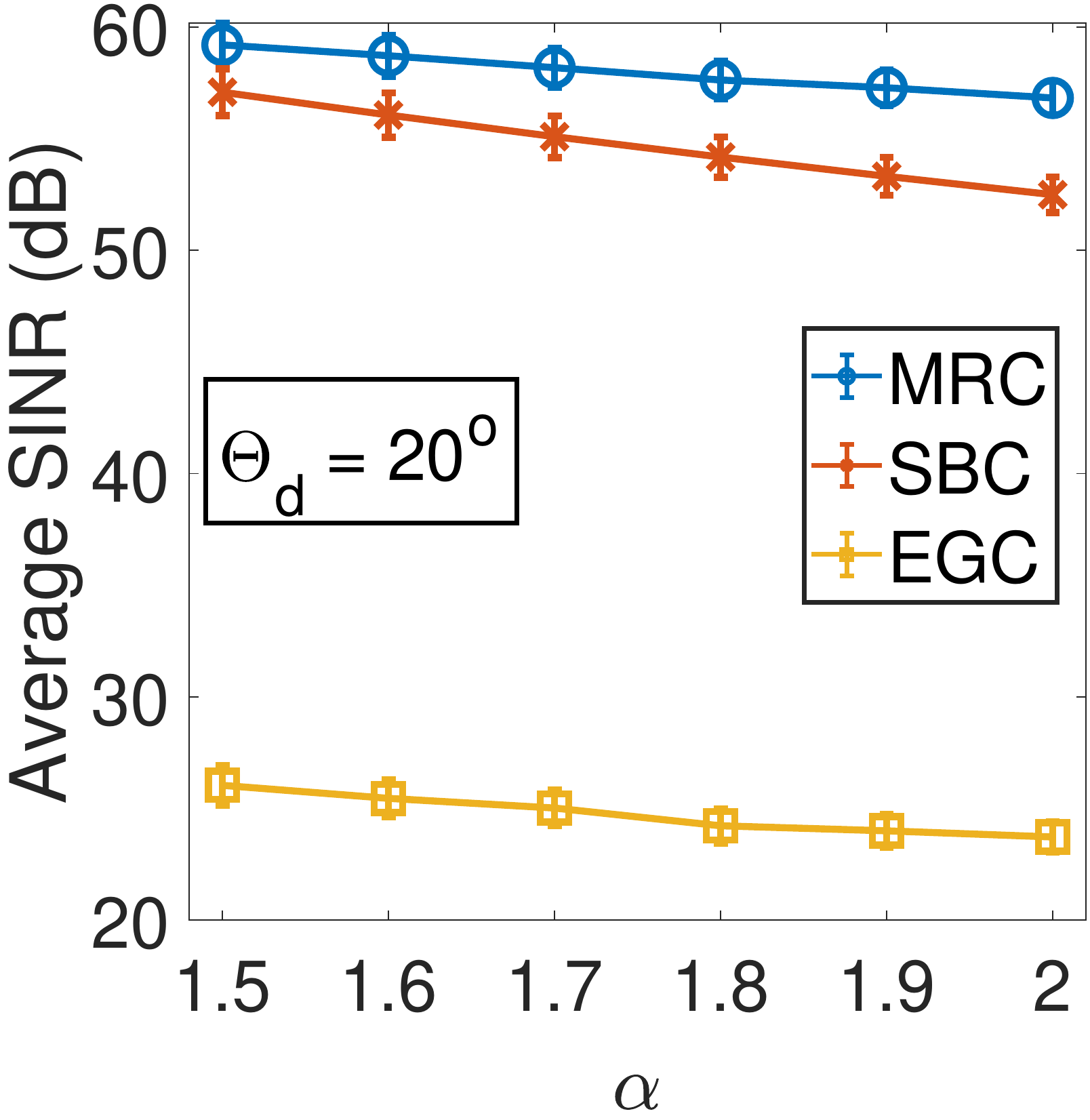}
		\vspace{-5pt}
		\subcaption{$\Theta_d = 20^o$}
		\label{fig:sinr_AL_20}
	\end{subfigure}\\
	\begin{subfigure}{.5\linewidth}
		\centering%
		\includegraphics[width=.75\linewidth]{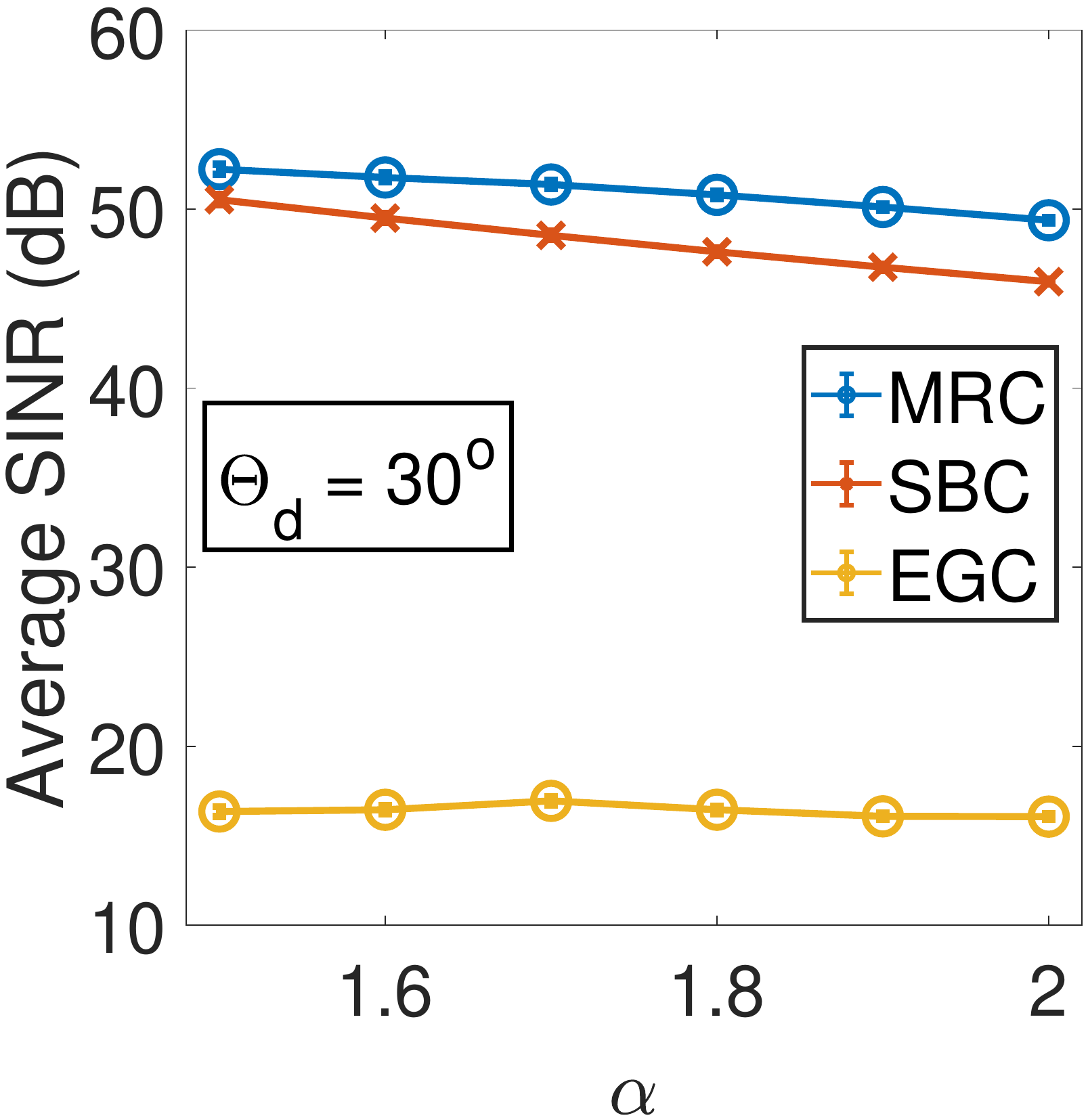}
		\vspace{-5pt}
		\subcaption{$\Theta_d = 30^o$}
		\label{fig:sinr_AL_30}
	\end{subfigure}%
	\begin{subfigure}{.5\linewidth}
		\centering%
		\includegraphics[width=.75\linewidth]{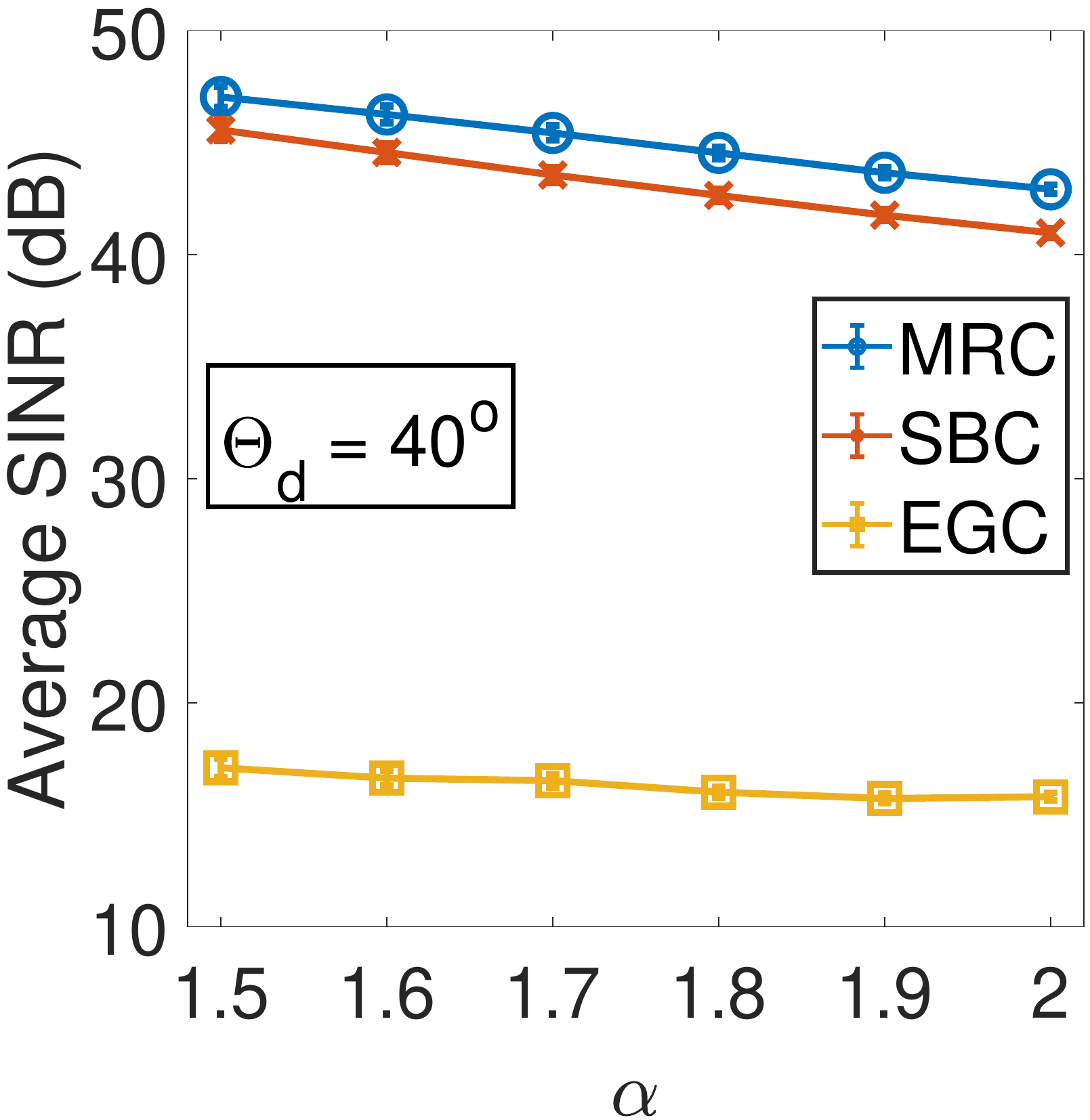}
		\vspace{-5pt}
		\subcaption{$\Theta_d = 40^o$}
		\label{fig:sinr_AL_40}
	\end{subfigure}%
	\vspace{-5pt}
	\caption{Average SINR achieved using different diversity combining techniques (with 95$\%$ confidence interval)}
	\label{fig:sinr_vs_alpha}
	\vspace{-10pt}
\end{figure}

We performed numerical analysis using $\alpha\ge\alpha_{min}$ for different values of angular distance $\Theta_d$ to determine $\alpha^{opt}$. We considered three different diversity combining techniques, EGC, MRC, and SBC for calculating the SINR at the VR-headset receiver. We considered 121 users located uniformly in the VR-arena. We observed in Section \ref{realdata} that the pitch and roll values stay within $-55^o$ to $55^o$ in $96.13\%$ and $99.43\%$ of the time respectively. So, in the simulations we considered roll and pitch values between $-60^o$ to $60^o$ and yaw values between $-180^o$ to $180^o$. Fig. \ref{fig:sinr_vs_alpha} displays the average SINR values for EGC, MRC, and SBC. We can observe that in all cases the SINR decreases with increase in $\alpha$ and the average SINR is maximum when $\alpha=1.5$. We can also see that for MRC and SBC, the SINR values are well above the required value of 20dB \cite{abari2017enabling} for VR applications. But, the achieved SINR values can be very poor for EGC since the signals received by all the PDs are added with equal weight.

\subsection{Appropriate number of photo-detectors}

\begin{figure}
	\centering
	\includegraphics[width=.75\linewidth]{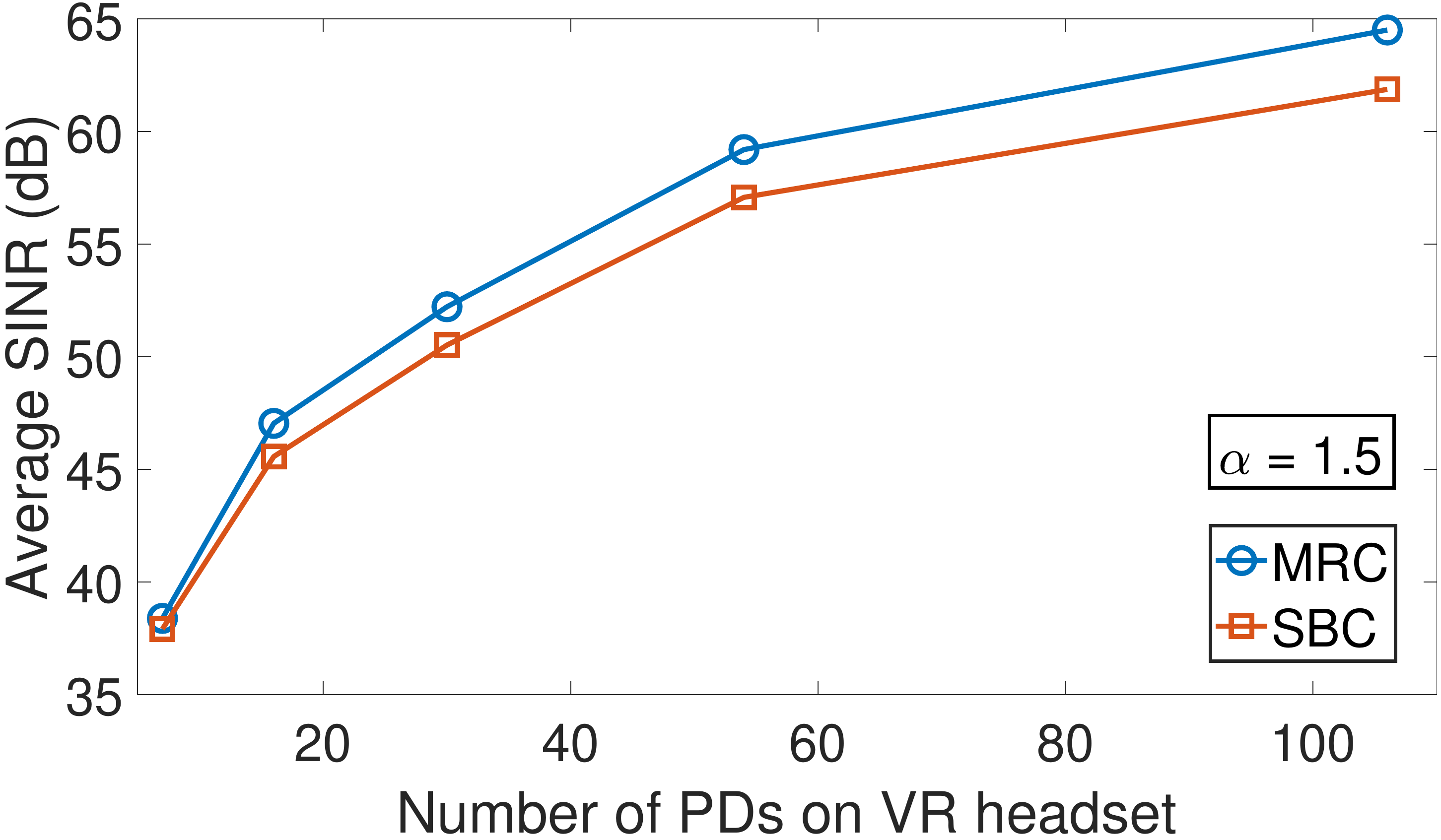}
	\caption{Higher SINR achieved using larger number of PDs}
	\label{fig:sinr_vs_Npd}
	\vspace{-20pt}
\end{figure}

In this section, we look at how the number of PDs on the headset relates to the SINR achieved using MRC and SBC techniques at the VR-headset receiver. We do not consider EGC because it cannot provide the required SINR of 20dB for VR applications. We calculate the number of detectors $N_{PD}$ from (\ref{eq:Npd}), considering $\theta_d = 15^o, 20^o, 30^o, 40^o, 60^o$, $\alpha_{min} = 1.5$, and FOV = $\alpha_{min}\times\theta_d$. We can observe from Fig. \ref{fig:sinr_vs_Npd} that, achieving higher SINR values requires larger number of PDs. For example, at least 20 PDs are required to attain SINR of 50dB and more than 100 PDs are required to achieve SINR of more than 60dB. We observe sufficient SINR$\ge$20dB even for smaller number of PDs since FOV is selected as $\alpha_{min}\times\theta_d$ to ensure $100\%$ connectivity for any head orientation.
	
As shown in Table \ref{tab:tab1}, we considered the LED modulation bandwidth B = 10MHz. Thus, the achieved SINR values yields data rates between 50-60 Mbps. As discussed earlier, recent literature \cite{hussein201625,denault2013efficient} has shown that RGB-LDs can generate white light that has comparable properties with white LED sources. The modulation bandwidth of such RGB-LDs can be multiple GHz and hence can provide multi-Gbps data rate while maintaining SINR $\ge$ 20dB for $\theta_d<60^o$.

\section{Conclusion}
\label{sec:conclusion}
\vspace{-5pt}
In this paper, we proposed the application of VLC for high quality video transfer in a VR-arena. Multiple LED/LD transmitters can be deployed in such an arena to establish optical attocells to provide service to multiple mobile VR-users. We presented the design of a multi-detector VR-headset that can provide optical link connectivity for any random head orientation. We provided simulation analysis on how an optimal placement of the PDs on the headset can help achieve the desired SINR values, thus high quality VR experience. We only considered the effect of the LOS component of the VLC channel in this work. We will consider the non-line-of-sight (NLOS) components in our future work. We plan to develop a proof-of-concept prototype of the proposed VR-headset and perform test-bed experiments to assess its performance.



\vspace{-5pt}
\bibliographystyle{IEEEtran}
\bibliography{references}

\end{document}